\documentclass[12pt]{article}
\usepackage{times}
\usepackage{geometry}
\usepackage[utf8]{inputenc}
\usepackage{enumitem,amssymb,graphicx}
\usepackage{ragged2e}
\usepackage{natbib}
\usepackage[dvipsnames]{xcolor}
\newlist{thematic}{itemize}{8}
\setlist[thematic]{label=$\square$}
\usepackage{pifont}
\newcommand{\aap}{A\&A}

\newcommand{\apjs}{ApJSS}

\newcommand{\apjl}{ApJ}

\begin{document}

\begin{center}
\textsc{Astro2020 Science White Paper} \\
\smallskip
\begin{large}
\textit{\textcolor{Mahogany}{\textbf{Physical Conditions in the Cold Gas of Local Galaxies}}} \\
\end{large}
\end{center}
\medskip
\noindent

\noindent \textbf{Thematic Areas:} \hspace*{60pt} $\square$ Planetary Systems \hspace*{10pt} $\checkmark$  \textbf{Star and Planet Formation} \hspace*{20pt}\linebreak
$\square$ Formation and Evolution of Compact Objects \hspace*{31pt} $\square$ Cosmology and Fundamental Physics \linebreak
  $\square$  Stars and Stellar Evolution \hspace*{1pt} $\square$ Resolved Stellar Populations and their Environments \hspace*{40pt} \linebreak
  $\checkmark$    \textbf{Galaxy Evolution}   \hspace*{45pt} $\square$             Multi-Messenger Astronomy and Astrophysics \hspace*{65pt} \linebreak
  
\noindent \textbf{Principal Authors:} Adam K. Leroy (Ohio State University, \texttt{leroy.42@osu.edu})

\medskip

\noindent \textbf{Co-authors:} Alberto D. Bolatto (University of Maryland) \\
Timothy A. Davis (Cardiff University) \\
Aaron S. Evans (University of Virginia) \\
Andrew Harris (University of Maryland) \\
Philip Hopkins (California Institute of Technology) \\
Annie Hughes (Institut de Recherche en Astrophysique et Planétologie) \\
Remy Indebetouw (University of Virginia, National Radio Astronomy Observatory) \\ 
Kelsey E. Johnson (University of Virginia) \\
Amanda A. Kepley (National Radio Astronomy Observatory) \\
Jin Koda (Stony Brook University) \\
David Meier (New Mexico Tech, National Radio Astronomy Observatory) \\
Eric Murphy (National Radio Astronomy Observatory) \\
Desika Narayanan (University of Florida) \\
Erik Rosolowsky (University of Alberta) \\
Eva Schinnerer (Max Planck Institute for Astronomy, National Radio Astronomy Observatory) \\
Jiayi Sun (Ohio State University) \\
Christine Wilson (McMaster University) \\
Tony Wong (University of Illinois Urbana-Champaign) \\
\medskip

\noindent \textbf{Abstract  (optional):} We describe a next major frontier in observational studies of galaxy evolution and star formation: linking the physical conditions in the cold, star-forming interstellar medium to host galaxy and local environment. The integrated gas content of galaxies has been surveyed extensively over the last decades. The link between environment and cold gas density, turbulence, excitation, dynamical state, and chemical makeup remain far less well understood. We know that these properties do vary dramatically and theoretical work posits a strong connection between the state of the gas, its ability to form stars, and the impact of stellar feedback. A next major step in the field will be to use sensitive cm-, mm-, and submm-wave spectroscopy and high resolution spectroscopic imaging to survey the state of cold gas across the whole local galaxy population. Such observations have pushed the capabilities of the current generation of telescopes. We highlight three critical elements for progress in the next decade: (1) robust support and aggressive development of ALMA, (2) the deployment of very large heterodyne receiver arrays on single dish telescopes, and (3) development of a new interferometric array that dramatically improves on current capabilities at cm- and mm-wavelengths ($\sim 1{-}115$~GHz).

\pagebreak

\begin{center}
\textcolor{Mahogany}{{\bf Motivation and Goals}}
\end{center}
\vspace{-3mm}

How, when, and where do galaxies form stars? Our best observational constraints on this fundamental question come from comparing the physical conditions in the cold, dense phase of the interstellar medium (ISM) to tracers of star formation activity and stellar feedback. Over the last decades we have made significant progress measuring star formation activity across redshift. Our progress quantifying the key physical conditions in the cold gas --- e.g., density, temperature, turbulence, gravitational boundedness --- has been slower. Fortunately, advances in instrumentation and theoretical modeling promise major progress in this area through the 2020s and beyond.

A major goal for the next decade should be a systematic approach to measuring the physical state of the cold ISM across the full range of galactic conditions and environments found in the local universe. Combining these new observations with measurements that trace recent star formation and stellar feedback will provide the foundation for \textbf{an observationally-vetted general theory of the physics of cold gas and star formation in galaxies---a key ingredient for modeling galaxy formation and evolution}. We highlight three major questions to be addressed:

\begin{enumerate}
\setlength\itemsep{0.25em}
\item How do the distributions of density, turbulence, temperature, and dynamical state in the cold ISM depend on host galaxy and local environment?

\item How do these physical conditions regulate the rate at which stars form from cold gas and the properties of the newborn stars and clusters? 

\item How do these physical conditions reflect and regulate the impact of stellar feedback?

\end{enumerate}

\begin{center}
\textcolor{Mahogany}{{\bf State of the Field}}
\end{center}
\vspace{-3mm}

Our knowledge of cold gas in galaxies has progressed dramatically in recent decades. Building on the pioneering work of \citet{KENNICUTT98A} and \citet{YOUNG95}, the relationships between star formation and the total molecular and atomic gas mass of galaxies have been explored at low and high redshift \citep[e.g., see][]{CARILLI13,TACCONI18}. Large surveys have measured the integrated cold gas masses and star formation rates of hundreds (for H$_2$) to thousands (for {\sc Hi}) of galaxies \citep[e.g.,][]{SAINTONGE17,HAYNES18}. CO and {\sc Hi} mapping surveys have quantified the resolved relationship between gas and star formation in hundreds of galaxies \citep[e.g.,][]{BOLATTO17}. Finally, surveys of the Milky Way have increasingly taken a ``top down'' view of gas and star formation \citep[e.g.,][]{LEE16,VUTISALCHAVUAKUL16}.

Meanwhile, theoretical work has explored the link between the physical state of the cold gas and star formation. There are now at least half a dozen analytic theories of star formation in turbulent clouds \citep[e.g.,][]{PADOAN14,BURKHART18}. A related body of work focuses on self-regulated star formation in galaxy disks \citep[e.g.,][]{OSTRIKER10,KRUMHOLZ18}. These models describe the long-term, large-scale equilibrium in the disk, as achieved by stellar feedback and potentially influenced by galactic dynamics and gas flows into galaxies \citep[e.g.,][]{SANCHEZALMEIDA14}. Numerical simulations of clouds, parts of galaxies, and individual galaxies have resolved the critical spatial scales for many feedback mechanisms and explored the processes of star formation and feedback in turbulent clouds \citep[e.g.,][]{WALCH15,PADOAN16}. 

In these theories, the rate of star formation in an individual molecular cloud depends on its physical properties, including its dynamical state and internal density distribution \citep[e.g.,][]{FEDERRATH12}. The structure of the cold gas on larger scales reflects an interplay of gravity, galactic dynamics, and stellar feedback. The details of stellar feedback, in turn, depend on the properties and locations of the cold gas relative to sources of feedback \citep[e.g.,][]{WALCH15}. \textbf{Thus, the physical conditions in the cold gas determine, and are sensitive to, the processes of star formation and stellar feedback.} 

Not only the overall rate, but also the nature of star formation appears to depend on cold gas properties. For example, the fraction of stars born in clusters has been shown to vary as a function of environment \citep[e.g.,][]{ADAMO15,JOHNSON16,GINSBURG18} while the stellar initial mass function appears ``bottom heavy'' in massive early type galaxies, which are thought to assemble in bursts of intense star formation at high redshift \citep[e.g.,][]{CONROY12,CAPPELLARI16}. \textbf{There is thus good evidence that the clustering -- and perhaps even the population -- of young stars depends on physical conditions in the cold gas.}

Studies within the Milky Way and nearby galaxies have found that the cold ISM exhibits a diversity of physical conditions. Densities range from a few 10s of particles per cm$^{-3}$ in cold {\sc Hi} and diffuse molecular gas, up to greater than $10^5$~cm$^{-3}$ in dense cores, filaments, galactic nuclei, and merging galaxies. Temperatures span from $\lesssim 10$~K in the coldest, densest parts of clouds, up to more than $100$~K in sites of recent star formation and starburst nuclei. Turbulent Mach numbers, inferred from highly supersonic line widths, increase from $\sim 10$ in Milky Way clouds to $>100$ in starbursts and galaxy centers. Reflecting the gamut of turbulence and gas density, the 
effective turbulent ``pressure'' in the cold ISM also varies dramatically. 

Due to technological limitations, our knowledge of how the properties of the cold ISM vary with physical conditions comes from detailed case studies that were focused on a few of the brightest and/or closest galaxies. These galaxies sample only a small range of environments and are not necessarily representative of the galaxy population as a whole.  Pursuing larger surveys, however, remains difficult with current instrumentation. {\bf Pushing detailed spectroscopy and imaging of cold gas into the big survey era should be a major goal for radio, mm-wave, and infrared astronomy moving into the next decade.} This observational effort will allow us to probe the full range of conditions across galaxy disks or the full galaxy population, and to link these properties to the star formation measured in these systems.

\begin{center}
\textcolor{Mahogany}{{\bf Key Goal \# 1 --- Detailed Spectroscopic Surveys of Cold Gas}}
\end{center}
\vspace{-3mm}

The density, excitation, and chemical composition of cold gas represent the next major frontier in observations of the cold ISM. Density plays \textit{the} central role in many theories of star formation. Excitation traces the impact of feedback on the gas and strongly affects the emissivity of the gas. As a result, excitation is crucial for interpreting observations at all redshifts because it relates closely to ``conversion factors,'' i.e., molecular mass-to-light ratios. Finally, we know that the abundances of molecular species vary in a way that reflects local conditions and feedback.

\textbf{Multi-line mm- and sub-mm wave spectroscopy represents the best tool to access density, excitation, and chemistry in the cold gas.}  Molecular rotational transitions  at cm, mm, and sub-mm wavelengths span a wide range of effective critical densities and excitation requirements. By observing multiple transitions with different density and excitation sensitivities, it is possible to constrain the distributions of density and temperature in the cold gas. Meanwhile, observing many transitions of the same species constrains the abundance of the molecule. 

To date, the main obstacle to wide-field spectroscopic mapping surveys has been the faintness of key molecular emission lines. For example, the HCN~(1-0), HCO$^+$~(1-0) and CS(2-1) lines are commonly used as ``dense gas tracers'' \citep[e.g.,][]{GAO04B,USERO15} because they are the brightest high-critical density $\lambda \sim$ 3--4~mm transitions. These lines are already 10--50 times fainter than the bright CO lines that trace the bulk of the molecular gas in galaxies, and were the focus of the last generation of multi-galaxy surveys. Analogous dense gas mapping surveys would require prohibitively large time requests ($\sim$100--2,500 more time) to observe large samples. This has pushed the field towards detailed studies of bright objects \citep[e.g.,][]{WATANABE14,MEIER15,AALTO15}, which have yielded important physical insights but not yet synthetic knowledge that describes the molecular gas reservoir across the galaxy population.

In the next decade(s), we need to overcome these technological obstacles and \textbf{move towards multi-transition, multi-species mapping of diverse samples of galaxies. This will allow us to quantitatively link density, excitation, and chemistry to galaxy structure and galaxy evolution across redshift.} Inferring the density distribution requires observing lines that span a wide range of effective critical densities. To constrain excitation, one observes multiple transitions of the same species. 
In practice this entails mix-and-matching among molecular species, making it critical to span enough chemical diversity to overcome uncertainties in abundance. Both of these tasks are easier if one can observe the optically thin transitions (usually from isotopologues), but these are typically a factor of $\approx$10 fainter again. Realistically, \textbf{the long-term requirement here is to survey large parts of the cm, mm, and sub-mm spectrum with sensitivity to lines 100$\times$ fainter than the bulk-gas tracing $^{12}$CO lines.} This sensitivity provides access to lines from many molecular species, with a wide range of densities and temperatures required for excitation \citep[e.g., see][]{MARTIN05,MEIER12,MEIER15,PETY17}. The abundances of these species trace a variety of conditions, e.g., the presence of shocks, AGN, CO freeze out, etc. In the intermediate to long term, surveying the full cm to sub-mm spectrum across normal disk and elliptical galaxies, starburst galaxies, dwarf galaxies, and merging galaxies will give us the same rich physical diagnostics currently accessible to optical surveys of ionized gas.

\begin{figure}[t!]
\vspace{-6mm}
\begin{center}
\hspace{0mm}\begin{minipage}[h]{0.575\linewidth}
\includegraphics[width=0.95\textwidth]{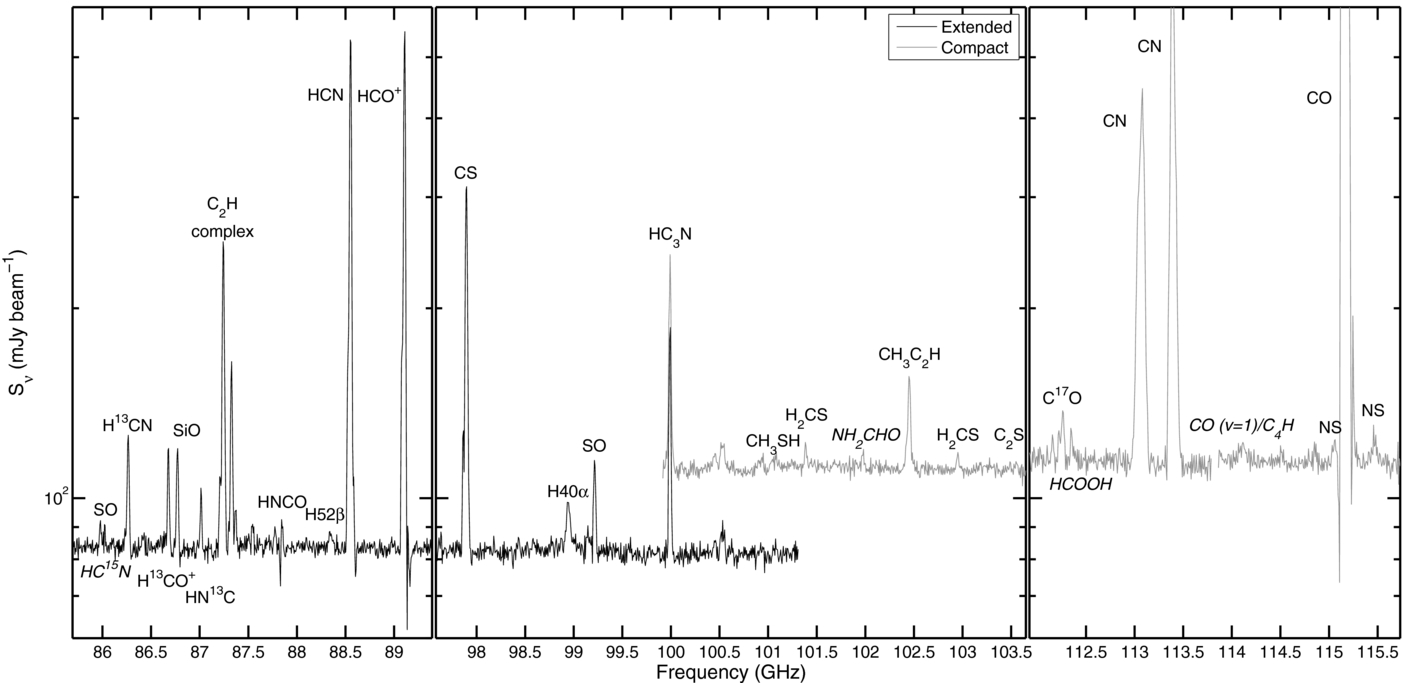}
\vspace{-3mm}\end{minipage}
\begin{minipage}[h]{0.40\linewidth}
\caption{\textit{Illustration of a part of the rich cm-, mm-, and sub-mm spectrum, here ALMA observations of the $\nu = 85{-}115$~GHz emission from the starburst nucleus of NGC~253 \citep{MEIER15}. The cm- to submm- spectrum is full of sensitive diagnostics of density, excitation, and interstellar chemistry.}}
\end{minipage}
\end{center}
\vspace{-7mm}
\end{figure}

\begin{center}
\textcolor{Mahogany}{{\bf Key Goal \# 2 --- High Resolution Spectroscopic Imaging of Cold Gas}}
\end{center}
\vspace{-3mm}

While spectroscopy yields the microphysical state of the gas, high resolution spectroscopic imaging traces the formation and evolution of clouds, cores, and filaments. This provides the immediate causal link between cold gas properties and galactic environment. The low-$J$ $^{12}$CO lines and 21-cm {\sc Hi} line trace the structure and kinematics of molecular and atomic gas. At scales of $100{-}500$ parsecs, these trace gas flows associated with arms, bars, and other dynamical features of the galaxy. At scales of $\sim 30{-}100$~pc, they trace the properties of individual clouds of cold gas. Observations at scales of $\sim 1{-}10$~pc access cloud substructure. And at scales of $\sim 0.1{-}1$~pc, filamentary structure, individual cores, and protoclusters become accessible.

In theory, molecular clouds form through a combination of large scale instabilities, conversion of atomic to molecular gas, and agglomeration of smaller structures \citep[e.g.,][]{DOBBS14}. Supersonic turbulence dominates their internal motion. Within clouds, a mixture of turbulence and external triggers collects low density gas into high density sub-structures. Stars form inside these dense sub-structures. Feedback from these newly formed stars then deposits energy and momentum back into the parent cloud, perhaps even destroying it.

Major uncertainties still surround each step of this process. What is the dominant formation mechanism for molecular clouds? What sets the mass function of molecular clouds and what drives observed variations with environment? How much molecular gas is bound as opposed to ``diffuse''? How, and on what scale, is turbulence driven? How much atomic gas is associated with cold, star-forming clouds? Both dynamical state (often expressed as the virial parameter) and mean density are expected to play a major role in star formation at multiple scales. \textbf{These predictions, however, remain weakly tested because the density, structure, and dynamical state of cold gas has not been observed at high resolution with good completeness across large parts of galaxies.}

\begin{figure}[t!]
\vspace{-6mm}
\begin{center}
\includegraphics[width=0.95\textwidth]{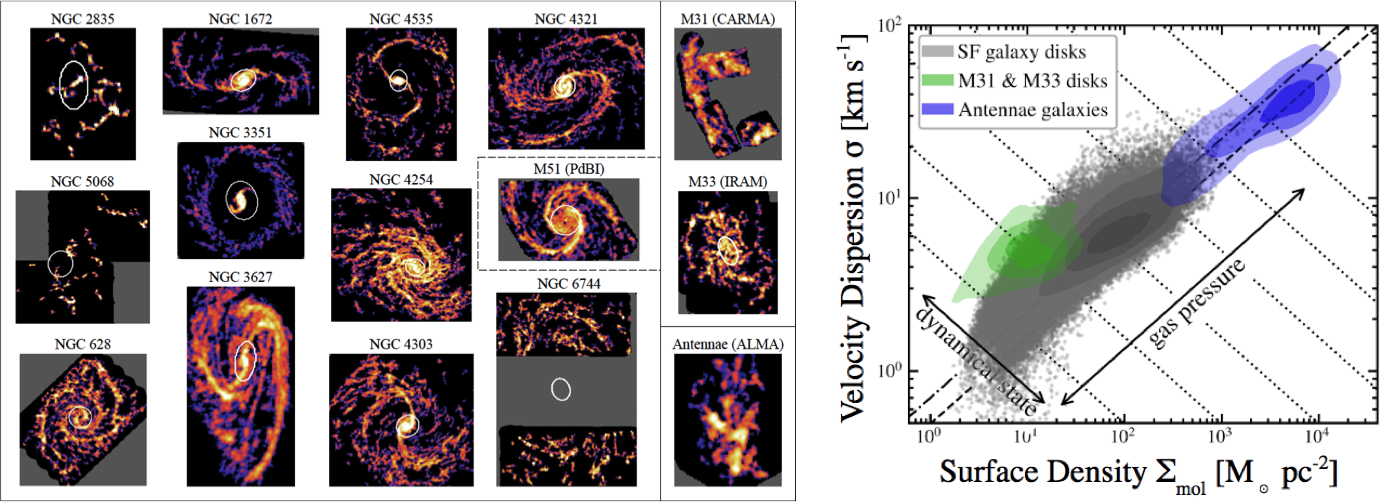}
\caption{\textit{High ($120$~pc) resolution CO maps of local galaxies, mostly CO~(2-1) from ALMA (left) and cloud-scale velocity dispersion vs. surface density for $\sim 30,000$ molecular cloud-sized beams \citep{SUN18}. Both plots show dramatic, systematic variation in the surface density, structure, line width and internal pressure of molecular clouds.}}
\end{center}
\vspace{-8mm}
\end{figure}

The key to progress here is \textbf{imaging the structure and kinematics of gas at scales from $<1{-}100$~pc across large parts of galaxies with the goal of measuring the density, organization, and dynamics of cold gas as a function of scale and environment.} Before now, such high resolution imaging has been deployed only in special environments or recovered only the brightest fraction of emission. As with the spectroscopy described above, the barrier has been sensitivity. The angular resolution $\lesssim 1''$ required to reach the scale of individual clouds in nearby normal star-forming disk galaxies can only be achieved by interferometers. Imaging spectral lines with high completeness (i.e., recovering a large fraction of the total flux), high fidelity, and high survey speed (allowing coverage of whole galaxies) requires very good surface brightness sensitivity. This requirement made galaxy-wide, high resolution CO surveys immensely expensive \citep[e.g.,][]{SCHINNERER13} and $\lesssim 1''$ 21-cm {\sc Hi} imaging has been almost impossible with current telescopes.

\begin{center}
\textcolor{Mahogany}{{\bf Relation to Current and Planned Facilities}}
\end{center}
\vspace{-3mm}

Moving into the 2020s, there are excellent prospects for obtaining a quantitative, detailed view of recent star formation, sources of stellar feedback, and the physical state of ionized gas. This opportunity is driven by an explosion in the power of optical integral field units (IFU), the legacy of the \textit{Hubble Space Telescope} in the local volume \citep[e.g.,][]{CALZETTI15,DALCANTON15}, and the upcoming launch of the \textit{James Webb Space Telescope}. The proposed \textit{Origin Space Telescope} and \textit{Lynx X-Ray Observatory} will also be exceptionally powerful probes of recent star formation and hot gas in galaxies. Following on from surveys like MANGA, CALIFA, and ATLAS$^{\rm 3D}$, the proposed fifth Sloan Digital Sky Survey includes IFU mapping of a key set of local star forming galaxies (the ``Local Volume Mapper''), which would provide an exquisite inventory of star formation and feedback across the local galaxy population.

It will be crucial to pair these data with observations measuring the physics of cold gas in galaxies at matching resolution. Because the vast majority of molecular gas is cold, the relevant photons have cm-, mm-, and sub-mm wavelengths. And because this emission is faint, carrying only a tiny amount of energy, its detection requires a large collecting area. As a result, our capabilities to date have been limited, but the situation is changing rapidly. Three concepts/facilities appear especially promising: (1) ALMA and its future development, (2) large format heterodyne arrays at single-dish facilities, and (3) major new interferometric arrays such as the proposed Next Generation Very Large Array (ngVLA).

\textbf{ALMA and Future Development:} ALMA is a capable facility, but even first-generation versions of the surveys described above still require a large time commitment. Making significant progress will take years and require continued robust support for the facility. From the current incarnation of ALMA, the next decade should yield CO maps with angular resolutions $\lesssim 1''$ ($\sim 100$~pc at the Virgo Cluster) for hundreds of galaxies, and spectroscopic data probing a suite of lines $\sim 10{-}20$ times fainter than CO for several dozens of galaxies. ALMA has outlined a development roadmap that would improve its capabilities in this area even further. The key developments are improved, larger bandwidth receivers (capable of simultaneously observing many molecular lines), the exploration of future focal-plane arrays, and perhaps the construction of a large sub-mm single dish telescope (e.g., ``AtLAST'') to pair with the array. These developments would multiply the speed of this type of survey by factors of a few. 

\textbf{Large Heterodyne Arrays:} Radio/submm cameras --- feed arrays or phased array feeds at lower frequencies --- multiply the survey speed of single dish telescopes like the Green Bank Telecope (GBT) and the Large Millimeter Telescope (LMT) by a factor similar to the number of pixels. The collecting areas of the GBT and the LMT already rival that of ALMA. Within the next decade, it should be possible to deploy massive heterodyne arrays with $>100$ elements, for example extending the ARGUS 3~mm array on the GBT. Such facilities will excel at mapping large areas with great surface brightness sensitivity and moderate resolution. This makes them ideal for spectroscopic surveys that resolve nearby galaxies but not individual clouds. They will also excel at accessing the detailed physical conditions in the cold gas of the Milky Way.

\textbf{A Next Generation Very Large Array:} Given the sensitivity limitations of ALMA at low ($\lesssim 115$~GHz) frequencies and the angular resolution limits of single-dish facilities, the next major step must be a telescope array that takes a significant leap forward in sensitivity beyond the current capabilities of the VLA and ALMA. The ngVLA concept represents an order-of-magnitude improvement over the VLA and ALMA at the critical frequencies for this science, $\sim 1{-}115$~GHz. The proposed sensitivity of the ngVLA would enable efficient sub-arcsecond imaging of density and excitation tracers, as well as parsec-scale mapping of nearby galaxies in CO. The gain in collecting area would also allow for $\lesssim 1''$ resolution imaging of the $\lambda=21$-cm {\sc Hi} transition, which would resolve the atomic gas associated with individual cold clouds in galaxies beyond the Local Group for the first time (the sensitivity of the current VLA limits the practical resolution of such imaging to $\sim 10''$). Although the Square Kilometer Array (SKA) will be a powerful instrument for {\sc Hi} imaging, neither phase of the SKA is designed to access the higher frequencies that are key to obtaining a complete picture of the physical state of the cold gas in local galaxies.


\pagebreak

\bibliography{scratch}

\end{document}